\documentclass[aps,prb,tightenlines,twocolumn,superscriptaddress]{revtex4-1}
\usepackage{amsmath}
\usepackage{amsfonts}
\usepackage{amssymb}
\usepackage{color}
\usepackage{hyperref}
\usepackage{dcolumn}% Align table columns on decimal point
\usepackage{bm}% bold math
\usepackage{slashed}
\usepackage[toc,page]{appendix}
\usepackage{float}
\usepackage{bbm}
\usepackage[pdftex]{graphicx}
\usepackage{color}

\renewcommand{\dag}{^{\dagger}}
\newcommand{\bk}{\mathbf{k}}
\newcommand{\sfk}{\mathsf{k}}
\newcommand{\bbt}{\mathbf{t}}
\newcommand{\bG}{\mathbf{G}}
\newcommand{\bt}{\mathbf{t}}
\newcommand{\br}{\mathbf{r}}
\newcommand{\bR}{\mathbf{R}}
\newcommand{\bx}{\mathbf{x}}
\newcommand{\cA}{\mathcal{A}}
\newcommand{\cC}{\mathcal{C}}
\newcommand{\cF}{\mathcal{F}}
\newcommand{\cK}{\mathcal{K}}
\newcommand{\cT}{\mathcal{T}}
\newcommand{\cP}{\mathcal{P}}
\newcommand{\bbZ}{\mathbbm{Z}}
\newcommand{\id}{\mathbbm{1}}
\newcommand{\ket}[1]{\left| #1 \right\rangle}
\newcommand{\bra}[1]{\left\langle #1 \right|}
\newcommand{\pars}[1]{\left( #1 \right)}
\newcommand{\brac}[1]{\left\{ #1 \right\}}
\newcommand{\eqn}[1]{\begin{equation} #1 \end{equation}}
\DeclareMathOperator{\dee}{d\!}
\DeclareMathOperator{\Tr}{Tr}

\begin{document}
\title{Bulk invariants and topological response in insulators and superconductors with nonsymmorphic symmetries}
\author{D{\'a}niel Varjas}
\affiliation{Department of Physics, University of California, Berkeley, California 94720, USA}
\author{Fernando de Juan}
\affiliation{Department of Physics, University of California, Berkeley, California 94720, USA}
\author{Yuan-Ming Lu}
\affiliation{Department of Physics, The Ohio State University, Columbus, Ohio 43210, USA}
\begin{abstract}
In this work we consider whether nonsymmorphic symmetries such as a glide plane can protect the existence of topological crystalline insulators and superconductors in three dimensions. In analogy to time-reversal symmetric insulators, we show that the presence of a glide gives rise to a quantized magnetoelectric polarizability, which we compute explicitly through the Chern-Simons 3-form of the bulk wave functions for a glide symmetric model. Our approach provides a measurable property for this insulator and naturally explains the connection with mirror symmetry protected insulators and the recently proposed $\bbZ_2$ index for this phase. More generally, we prove that the magnetoelectric polarizability becomes quantized with any orientation-reversing space group symmetry. We also construct analogous examples of glide protected topological crystalline superconductors in classes D and C and discuss how bulk invariants are related to quantized surface thermal-Hall and spin-Hall responses.
\end{abstract}
\bibliographystyle{apsrev4-1}

\maketitle

\section{Introduction}

The last decade has seen a major breakthrough in the search of novel phases of matter with the discovery of topological insulators and superconductors\cite{Hasan2010,Hasan2011,Qi2011}. The original predictions of these systems have already led to many experimental realizations, in a very fruitful endeavor that continues today. The key insight underlying this discovery is that the presence of a symmetry, in this case time-reversal symmetry ($\cT$), allows to define a new bulk topological invariant of the Bloch wave functions in the Brillouin Zone (BZ). In a gapped fermion system, this invariant cannot change unless the gap closes, defining a robust phase and protecting the existence of gapless boundary states. It was soon realized that other global symmetries in the Altland-Zirnbauer (AZ) classes\cite{AZ}, such as charge conjugation ($\cC$) and chiral symmetry, also give rise to new phases, leading to the periodic table\cite{Schnyder2008,Kitaev} of topological insulators and superconductors.

The classification based on global symmetries then lead to the natural question of whether lattice symmetries can give rise to new topological phases of matter. For example, it was realized early on that in the presence of lattice translations, one may define extra topological invariants in lower dimensional slices of the Brillouin Zone~\cite{BalentsMoore}, which lead to the concept of weak topological insulators \cite{Ran,Ran2010}. Point group symmetries such as rotations or reflections were also used to define new topological invariants and phases of matter, which were termed topological crystalline insulators\cite{Fu2011,AndoFu} (TCI) and superconductors (TCSC). Recent efforts in the field \cite{FuKaneInversion,BernevigPGS,BernevigPGS2,TeoHughesPGS,Zaanen,Chiu,Morimoto,Shiozaki,Lu,Mong} have been devoted to classifing these phases of matter protected by lattice symmetries in addition to global ones. Most of these previous works focused on symmorphic space groups for simplicity, i.e. groups where the full group is a semidirect product of the translation part and the point group. However, in view of the strong constraints that non-symmorphic symmetries place on the Bloch wave functions, one may expect that these symmetries can lead to richer structures, an idea that has drawn a lot of attention recently \cite{SidNS,FuGlide,ShiozakiNS,Mong,LiuNS,WatanabeNS,WatanabeNS2,WangLiuNSSC}. The question we consider in this paper is whether non-symmorphic symmetries, in particular glide reflections, can define a new class of topological insulators or superconductors. We will focus on the three dimensional case without time-reversal symmetry, where it has been predicted that a new TCI protected by glide symmetry indeed exists\cite{FuGlide}.

While in this work we will present explicit computations of microscopic topological invariants, our main conclusions can also be understood in a simple way by considering topological bulk responses. It is well known that a three dimensional topological insulator can be characterized by a quantized bulk electromagnetic response term of the type\cite{Qi2008}
\begin{equation}
S = \frac{\theta e^2}{16\pi h} \int d^4\mathsf{x}\; \epsilon^{\mu\nu\lambda\gamma} F_{\mu\nu} F_{\lambda\gamma} = \frac{\theta e^2}{4 \pi h} \int F \wedge F
\label{theta1}
\end{equation}
which is known as the magnetoelectric response (because $F \wedge F\propto \mathbf{E\cdot B}$) or the ``axion'' Lagrangian. The second equality is expressed in coordinate-free notation, which we will use from now on (see Appendix \ref{CS}). The magnetoelectric coupling $\theta$ is defined modulo $2\pi$, and the presence of time-reversal symmetry requires that $\theta = -\theta$. This implies that $\theta =0$ or $\pi$, and the second case corresponds to a strong topological insulator. $\theta$ can be computed microscopically from the Chern-Simons 3-form of the Berry connection, establishing a direct correspondence with the $\bbZ_2$ index. A physical consequence of $\theta = \pi$ is the presence of an odd number of massless Dirac fermions on the surface which are protected by time-reversal symmetry.

The same line of reasoning\cite{Mong} implies that the magnetoelectric coupling is quantized in the presence of any symmetry that sends $\theta\rightarrow -\theta$, such as mirror reflection\cite{FuMirror,Chiu} which reverses one spatial coordinate. In the presence of a surface that respects this symmetry, one must also have an odd number of Dirac cones. A three dimensional topological insulator can therefore be protected by either time-reversal or mirror symmetry. The magnetoelectric coupling in the second case can be computed microscopically from mirror Chern numbers\cite{FuMirror} at invariant planes.

The main result of this work is that glide symmetry also gives rise to quantized magnetoelectric coupling, which exactly corresponds to the $\bbZ_2$ invariant previously defined for glide protected topological crystalline insulators\cite{FuGlide}. This result can be simply rationalized by the fact the magnetoelectric coupling is a bulk response property which makes no reference to the lattice. Since a glide differs from a regular mirror only by a half translation, from the perspective of the bulk response both symmetries guarantee the quantization of the $\theta$ term, giving rise to a topological insulator when $\theta=\pi$. This result is explicitly proven in Appendix \ref{sec:trfproof} where we show in general that $\theta$ becomes quantized in the presence of any orientation-reversing space group symmetry.

In this work we also illustrate this general result by an explicit calculation of $\theta$ via the Chern-Simons form for a particular model of a glide symmetric TI, confirming the presence of a single Dirac fermion in the surface spectrum when $\theta=\pi$. In the second part of this work, we explain how these ideas can be naturally extended to superconductors without time-reversal symmetry in classes D and C. In these classes there are analogs of the magnetoelectric coupling for the thermal and spin response, respectively, and these can also be quantized in the presence of a glide symmetry. We will also present explicit models for these classes, and show that microscopic computations of the bulk topological invariants are consistent with the surface spectra.

\section{Magnetoelectric Coupling and $\mathbbm{Z}_2$ Invariant In Mirror and Glide Symmetric  Insulators}

The quantized magnetoelectric response in Eq.~\ref{theta1} has been long known to be the distinguishing feature of strong topological insulators\cite{Essin2009, Qi2008} with time-reversal invariance. The quantized coefficient in the action is known as the magnetoelectric polarization $\theta$, and is given by the formula
\eqn{
\label{eqn:CS3}
\theta = \frac{1}{4 \pi} \int_{BZ} \Tr\pars{\cA\wedge\dee\cA-\frac{2i}{3}\cA\wedge\cA\wedge\cA}.
}
where
$\cA^{nm} = i\bra{u^n}\dee\ket{u^m}$ for $n$ and $m$ conduction bands, and the integral is only gauge invariant modulo $2\pi$, consistent with the ambiguity in Eq.~\ref{theta1}. While $\theta$ is computed in this manner for translationally invariant systems, the corresponding topological response is robust against disorder that preserves the symmetry on average and can be defined in the presence of interactions. The quantization of $\theta$ can also be protected by spatial symmetries like inversion symmetry \cite{Turner2012,BernevigInv} and improper rotations\cite{BernevigPGS}, but cases where it is not possible to find a surface that preserves the symmetry lack protected gapless surface states.

The existence of a quantized magnetoelectric response in a bulk material has an important implication for the response of surface states. In the presence of a small perturbation that breaks time-reversal symmetry, the surface Dirac fermion becomes gapped, giving rise to a half-integer quantized Hall conductance\cite{Qi2008}. This cannot happen in a pure two-dimensional system without topological order and reflects the topological nature of the 3d bulk. This behavior can be understood by considering the surface as an interface between the bulk and the vacuum where $\theta$ changes from $\pi$ to 0. The surface can thus be modeled by a spatially dependent $\theta$
\begin{equation}
S = \frac{e^2}{4\pi h} \int \theta(x) F \wedge F.
\end{equation}
such that $\theta(x<0) = \pi$ and $\theta(x>0) = 0$ for an interface between the topological and trivial regions. Since $F \wedge F$ is a total derivative, one may integrate by parts to find the effective 2+1-D surface action of the Abelian Chern-Simons form
\begin{equation}
S = \Delta\theta \frac{e^2}{4\pi h} \int  A \wedge \dee A
\label{u1}
\end{equation}
which implies that the effective Hall conductance of the surface, given by $\Delta \theta = \pi$ in units of $e^2/(2\pi h)$, is a half-integer value. This half-quantized topological response can serve as an additional feature to distinguish a topological phase, and as we will see, can be generalized to other types of responses.

To determine the coefficient $\theta$ and identify a topological phase, one needs to explicitly evaluate Eq.~\ref{eqn:CS3}. The coefficient will be quantized in the presence of any symmetry that takes $\theta \rightarrow -\theta$, but the computation may simplify in different ways for different symmetries. For example, in the presence of both time-reversal and inversion symmetry, Eq.~\ref{eqn:CS3} can be related to the eigenvalues of the inversion operator at time-reversal invariant momenta\cite{FuKaneInversion}. In the presence of a mirror symmetry, the computation can be related to the mirror Chern numbers at mirror invariant planes. In Appendix~\ref{sec:trfproof} we prove the quantization of $\theta$  in the presence of a generic orientation-reversing space group symmetry with no reference to any of these simplifying circumstances.

The purpose of this section is to demonstrate the robust quantization of $\theta$ imposed by mirror and glide symmetries using microscopic tight-binding models. Starting with a brief review of crystalline insulators in class A with mirror symmetry, we show that the quantized magnetoelectric response can be obtained as the parity of the the integer-valued ($\bbZ$) topological invariant protected by the mirror. However, when the mirror symmetry is replaced by a glide\cite{FuGlide,ShiozakiNS}, \emph{only} a $\bbZ_2$ invariant survives, which corresponds to the quantized $\theta=0,\pi$.

\subsection{Mirror symmetry}
\label{AM}

In three dimensional insulators with mirror symmetry, topological invariants can be defined by considering the mirror invariant planes in the BZ, where bands have a definite mirror parity. Total and mirror Chern numbers can be thus defined for these invariant planes~\cite{FuMirror,hsieh}. The total Chern numbers for cuts perpendicular to the mirror plane vanish by symmetry, and for simplicity we assume they are zero in the mirror-invariant planes as well\footnote{In a gapped 3d insulator, the total Chern number in any parallel cut in momentum space must be the same integer, which is a 2d weak index. We omit this extra factor of $\bbZ$ in our further classification.}. In a system with mirror symmetry reflecting the $z$ axis, the bulk BZ has two mirror invariant planes\footnote{Note that not all space groups have two mirror-invariant $\bk$-planes. For example a tetragonal crystal only has one pointwise invariant plane with respect to a diagonal mirror. While this might affect the detailed classification with mirror symmetry, the conclusions about the quantization of $\theta$ remain valid.} at $k_z=0$ and $\pi$ (Fig.~\ref{glideBZ0} (a)). This allows us to label bands in these two planes by their mirror eigenvalue $\pm i^F$ ($F=0$ and $1$ for spinless and spinful fermions respectively), as no terms mixing the two sectors are allowed by symmetry. Chern numbers $C^{\pm}_{k_z}$ for the even and odd occupied bands are separately well defined for $k_z = 0$ and $\pi$. The mirror Chern number for a mirror-invariant plane $(k_z=0,\pi)$ is defined as the difference between the two sectors $C^M_{k_z} = \frac{1}{2}\pars{C^+_{k_z} - C^-_{k_z}}$. Consider, for example the case with nonzero Chern numbers for the even and odd sectors in the $k_z=0$ plane ($C^{\pm}_0 =\pm 1$) and vanishing Chern number for both sectors at $k_z = \pi$ ($C^{\pm}_{\pi} =0$), now $C^M_{0} = 1$ and $C^M_{\pi} = 0$.

A minimal Hamiltonian implementing this phase can be obtained starting from the 4-band model of a 3d TI with time-reversal symmetry\cite{BernevigPGS,Qi2008} $\cT = i \sigma_y\cK$:
\eqn{
\label{eqn:HAM}
H^{AM}_{\bk} = t_x\sin k_x \tau_y + t_y \sin k_y \sigma_z \tau_x + t_z \sin k_z \sigma_y \tau_x + m_{\bk} \tau_z
}
with $m_{\bk} = m - \sum_{\mu} \cos k_{\mu}$. This model is in the strong TI phase for $1<|m|<3$ with $\theta = \pi$. Now we can remove the time-reversal symmetry constraint and instead demand invariance under mirror $M = i^F \sigma_z$ reflecting the $z$ direction. The Hamiltonian (\ref{eqn:HAM}) formally has both symmetries, but the mirror allows different perturbations than time-reversal. It is easy to see that the two mirror sectors (with opposite $\sigma_z$ eigenvalues) now have opposite Chern numbers in the $k_z = 0$ plane and vanishing Chern numbers at $k_z =\pi$. Similar to the weak indices in time-reversal invariant insulators,  $C_0^M$ and $C^M_{\pi}$ separately rely on translational symmetry along $z$ direction but the \emph{strong} mirror Chern number $C^M_s = C^M_0 + C^M_{\pi}$ is robust against translational symmetry breaking, as long as mirror symmetry is preserved. We note that, with vanishing total Chern number (\textit{i.e.}~$C^+_{k_z}+C^-_{k_z}=0$), the quantized magnetoelectric coupling $\theta$ is completely determined by the strong mirror Chern number, $\theta = \pi C^M_s \pmod{2\pi}$. This is seen by counting surface modes: in the invariant planes $C^+_{k_z}=-C^-_{k_z}$ counts the number of chiral modes on the surface at $k_z=0$ or $\pi$ propagating right (left) in mirror sector $+$ ($-$). Each pair of counterpropagating modes forms a surface Dirac-cone, so if the total number is odd, the bulk has nontrivial $\theta$.

\subsection{Glide symmetry}

\begin{figure}[t]
\begin{center}
\includegraphics[width=8.5cm]{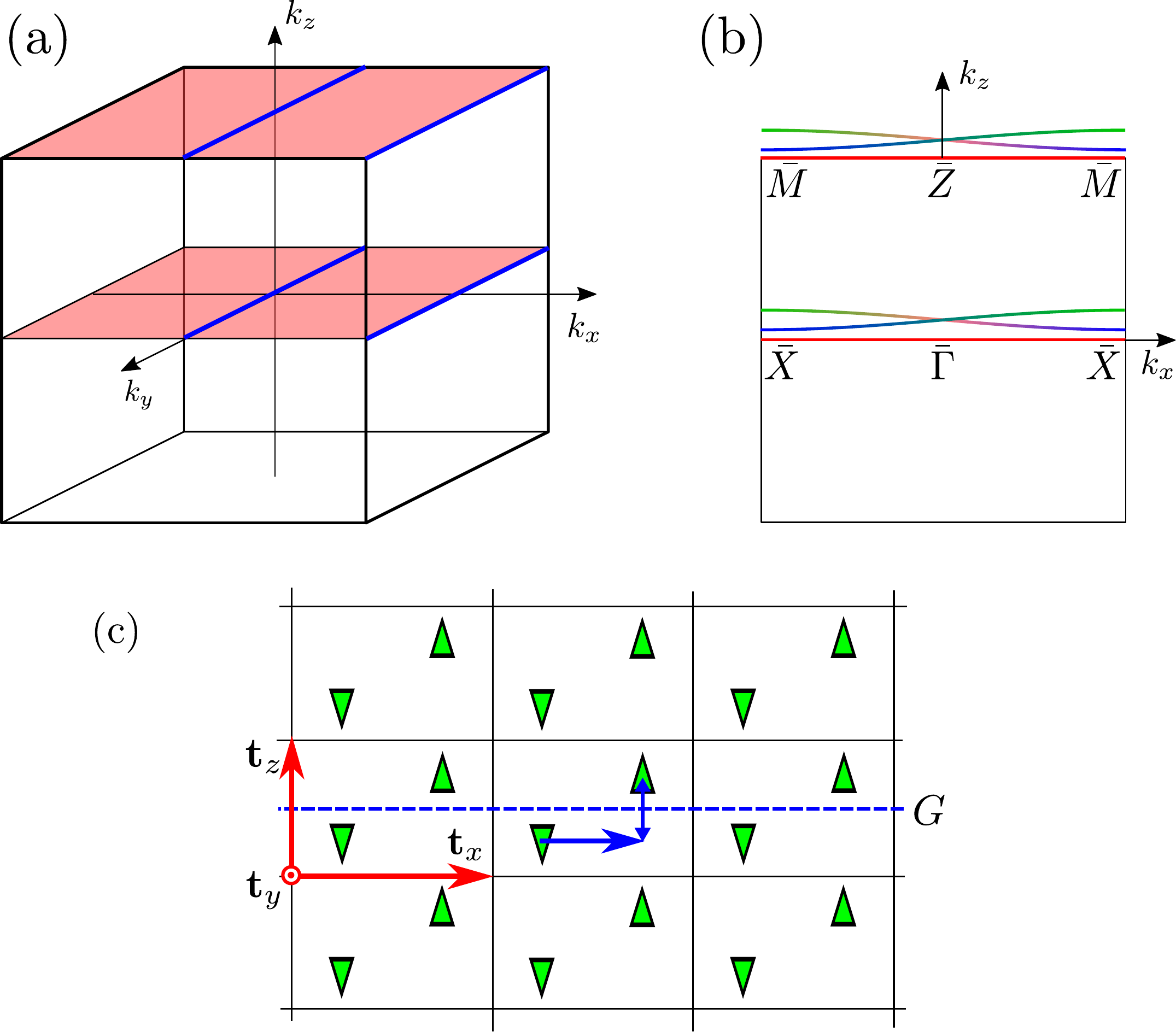}
\caption{(a) Bulk BZ of a glide or mirror symmetric crystal with the two invariant planes BZ in red (shaded) and invariant lines in blue (thick) if particle-hole symmetry is also present. (b) Surface BZ for a cut normal to $y$ with invariant lines in red (thick), labels for high symmetry points and a sketch of an occupied band pair along these lines. On the invariant planes ($k_z = 0$ and $k_z = \pi$) the glide eigenvalue (color code) distinguishes the two bands. (c) Example of a crystal with glide symmetry $G$ that reflects the $z$ direction and translates by half of the unit cell in the $x$ direction, the glide operator exchanges the two sublattices ($\bigtriangleup$ and $\bigtriangledown$). This pattern is repeated in parallell planes shifted perpendicular to the plane of the drawing.}\label{glideBZ0}
\end{center}
\end{figure}

In this section we present an alternate picture to understand the glide-protected $\mathbbm{Z}_2$ invariant first proposed in Ref~\onlinecite{FuGlide} (see also Appendix \ref{bentBZ}), in terms of the quantized magnetoelectric polarization. Let the glide $G = \brac{M_z| \bbt_x/2}$ reflect the $z$ direction and translate along $x$ by half of a unit cell (Fig.~\ref{glideBZ0}). $G^2 = \brac{(-1)^F \mathbbm{1}| \bbt_{x}}$ is a pure translation, with the sign depending on how a $2\pi$ rotation is represented. The eigenvalues of the glide operator are $\pm i^F e^{i k_{x}/2}$. As we traverse the Brillouin Zone (BZ) in the $x$ direction on a line that is pointwise invariant under the symmetry, the eigenvalues wind into each other. In a system respecting this symmetry if we follow a band with the $+i^F e^{i k_{x}/2}$ eigenvalue, it is connected to band with $-i^F e^{i k_{x}/2}$ at the zone edge, so the boundary condition for the Bloch wave functions is constrained to $\ket{u_{\bk+\bG_{x}}^+} \propto \ket{u_{\bk}^{-}}$ where $\bG_{x}$ is the reciprocal lattice vector parallel to $\bbt_{x}$. So bands in the presence of glide symmetry come in pairs that cannot be separated by a gap, as the crossings are protected by the fact that the pairs have different eigenvalues under the symmetry. The Chern number for each single band is ill defined as one band must evolve into the other on the zone boundary, only the total Chern number for the pair is a topological invariant. 

Again, for simplicity, we assume that the total Chern number for conduction/valence bands in any 2d cut parallel to the mirror plane vanishes. Nonzero values for perpendicular cuts are forbidden by mirror symmetry. The minimal model realizing the nontrivial phase\cite{FuGlide} is analogous to the mirror-symmetric case i.e. Eq. (\ref{eqn:HAM})
\begin{align}
\label{eqn:HAG}
H^{AG}_{\bk} &= t_x \sin\pars{\frac{k_x+\phi}{2}} \rho_x\tau_x +t_y \sin k_y \tau_y + \nonumber\\
&+ t_z \sin k_z \rho_z \tau_x + m_{\bk} \tau_z
\end{align}
with the glide operator $G_{\bk} = i^F e^{i k_x/2} \rho_x$. Here $m_{\bk}$ has the same form as in (\ref{eqn:HAM}). It is easy to see that the model is gapped for appropriate choice of parameters and band degeneracies can be removed almost everywhere in the BZ with symmetry allowed terms. Pauli matrices $\tau$ and $\rho$ act on orbital and sublattice degrees of freedom respectively and the Hamiltonian preserves glide symmetry $G_{\bk} H_{\bk} = H_{M_z \bk} G_{\bk}$ with $M_z \pars{k_x,k_y,k_z} =\pars{k_x,k_y,-k_z}$. We used the convention where operators and Bloch wave functions are not periodic in the BZ (see Appendix \ref{conv}), and the model can be regarded as either spinless ($F=0$) or spinful ($F=1$) with spin-polarized electrons, such that trivial bands with opposite $z$-component spin $S^z$ are pushed far over or below the Fermi level and can be omitted (note that $M_z$ is diagonal in the $S^z$ basis).

\begin{figure}[h]
\begin{center}
\includegraphics[width=8.5cm]{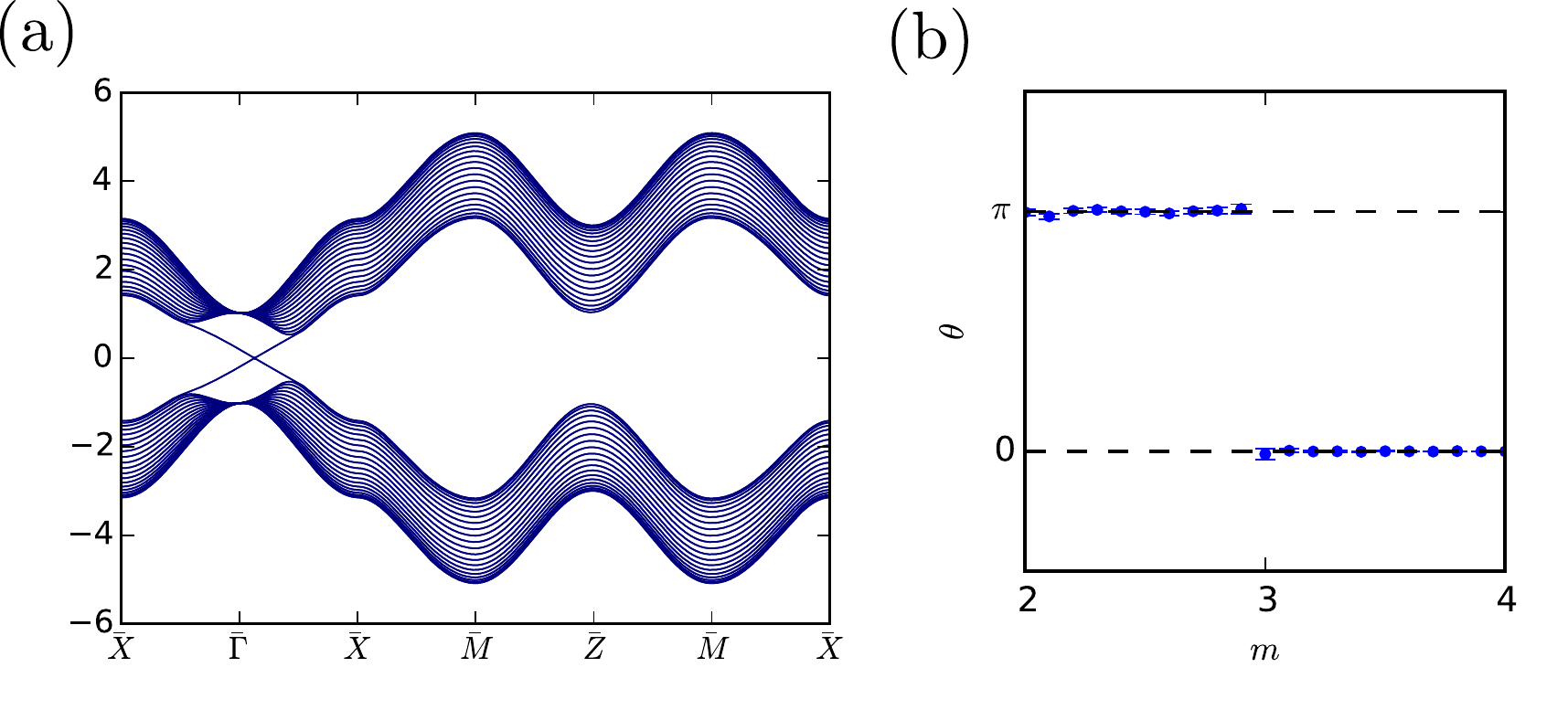}
\caption{(a) Band structure in slab geometry with 20 unit cells in the $y$ direction for glide protected class A topological insulator. We use $H^{AG}$ with $t_{\mu} = 1$, $m = 2$ and $\phi = 0.4$. The surface Dirac cone is at a generic momentum on the high symmetry line, the left and right moving branches are distinguished by their glide eigenvalues and cannot gap out. (b)~Evolution of $\theta$ from numerics while tuning across the transition from the topological to the trivial phase without breaking the glide symmetry. The error bars indicate two standard deviations (95\% confidence interval) of the Monte Carlo estimates.}\label{fig:ClassA}
\end{center}
\end{figure}

Regardless of the microscopic differences, in the macroscopic translational-invariant electromagnetic response theory there should be no distinction between a mirror and a glide, and, as we formally prove in Appendix~\ref{sec:trfproof}, $\theta$ is quantized as $0$ or $\pi$ just like with a mirror. We numerically\cite{LepageVegas} verified that in the nontrivial phase of this model, $\theta = \pi$ is robust against symmetry preserving perturbations (Fig.~\ref{fig:ClassA}, for details see Appendix \ref{CS}). We would like to point out that starting from a mirror symmetric TCI (for example (\ref{eqn:HAM})) one can double the unit cell and weakly break down separate mirror and half translation symmetries to a glide without closing the gap or changing the value of $\theta$. While the mirror Chern numbers are no longer well defined, the $\bbZ_2$ invariant defined by $\theta$ survives. We emphasize that $\theta$ is a macroscopic response quantized by macroscopic mirror symmetry, so it is robust against symmetry preserving interactions and disorders which preserve the symmetry on average\cite{Fulga2012}, both in the case of glide and mirror symmetry. To summarize, we provided a physical understanding of the $\mathbbm{Z}_2$ invariant introduced in Ref.~\onlinecite{FuGlide} in terms of quantized magnetoelectric coupling.

\section{Topological Crystalline Superconductors in Class D}

In the previous section we have shown how the magnetoelectric coupling in insulators can be quantized in the presence of symmetries other than $\cT$, in particular a mirror or a glide. We now show how these considerations can also be applied to superconductors in three dimensions where $\cT$ is broken. In this section we consider a superconductor with no other local symmetry but the particle-hole symmetry of the Bogoliubov-de Gennes Hamiltonian, which belongs to class D. We will show that the presence of an additional glide symmetry protects the existence of a topological crystalline superconductor with a single Majorana cone at the surface. In section \ref{sec:classC} we will consider the analogous problem in the presence of SU(2) spin rotation symmetry.

\subsection{Bulk invariant and surface thermal Hall conductance}

The reason why a glide can protect a topological superconductor in three dimensions without time reversal symmetry is that this phase is also characterized by $\theta$ term that is quantized with any orientation-reversing symmetry. The reasoning is analogous to the one used for insulators. We first consider the case of a superconductor with time-reversal symmetry, in class DIII. In three dimensions, class DIII has an integer topological invariant $\nu\in \bbZ$ which counts the number of Majorana cones at the surface. In the same way as the insulator, in the presence of a weak perturbation that breaks time reversal symmetry, the surface becomes gapped and each Majorana cone contributes half of the minimal thermal Hall conductance of a 2d superconductor (that is half of the minimal value for a 2d insulator)\cite{Kane1996,Ryu2012}:
\eqn{
\frac{\kappa_{xy}}{T} = \frac{(\pi k_B)^2}{3h} \frac{\nu}{4}.
}

Formally, a class D superconductor is the same as an insulator with an extra antiunitary particle-hole symmetry $\cC$ that anticommutes with the Hamiltonian and squares to $+\id$, because a particle-hole symmetric Bloch Hamiltonian for insulators has the same form as the Bogoliubov-de Gennes (BdG) Hamiltonian for superconductors. The only important difference is that only half of the degrees of freedom in the BdG Hamiltonian are physical, since all negative energy states correspond to the annihilation operators of the positive energy excitations over the BdG ground state. This is why the surface Dirac cones in insulator case reduce to surface Majorana cones in the superconductor case.

The fact that the insulator and superconductor problems are formally the same allows us to use the BdG Hamiltonian the same way as the Bloch Hamiltonian to calculate $\theta$ from the band structure of a glide-symmetric superconductor, which must be quantized to $0,\pi$ by the same reason as in the insulator case. When $\theta=\pi$, this implies an odd number of surface Majorana cones, and a half-integer thermal Hall conductance when the glide symmetry is broken. This is not allowed in a purely two-dimensional gapped superconductor with no ground state degeneracy, where $\kappa_{xy}/T$ is always quantized in integer multiples of $(\pi k_B)^2/(6h)$.

\subsection{Microscopic model with glide plane in 3d}

The explicit model for a glide symmetric superconductor in class D is very similar to the insulator in Eq.~(\ref{eqn:HAG}), but with an extra particle-hole symmetry $\cC^2 = \id$. Again we consider a glide plane reflecting the $z$ direction and translating along $x$, $G = \brac{M_z| \bt_x/2}$. We represent charge conjugation as $\mathcal{C} = \tau_x \mathcal{K}$ where $\mathcal{K}$ is the complex conjugation operator and the $\tau$ act on the particle-hole degree of freedom. As charge conjugation acts locally, it has to commute with the glide as $G_{-\bk} \cC = \cC G_{\bk}$. The Hamiltonian (\ref{eqn:HAG}) with $\phi=0$ possesses these symmetries for spinless fermions, however, such systems do not naturally appear as superconductors and require fine tuning as insulators. For the rest of this section we assume the physical case of spinful fermions, for which the appropriate choice of the glide representation is $G_{\bk} = i e^{i k_x/2} \rho_x \tau_z$ (see Appendix \ref{BdG} for details about symmetry representations in BdG systems). A simple Hamiltonian respecting these symmetries is:
\eqn{
\label{eqn:HDG}
H^{DG}_{\bk}= t_x \cos\frac{k_x}{2} \rho_y\tau_x +t_y \sin k_y \rho_z\tau_x + t_z \sin k_z \tau_y + m_{\bk} \tau_z.
}

\begin{figure}[h]
\begin{center}
\includegraphics[width=8.5cm]{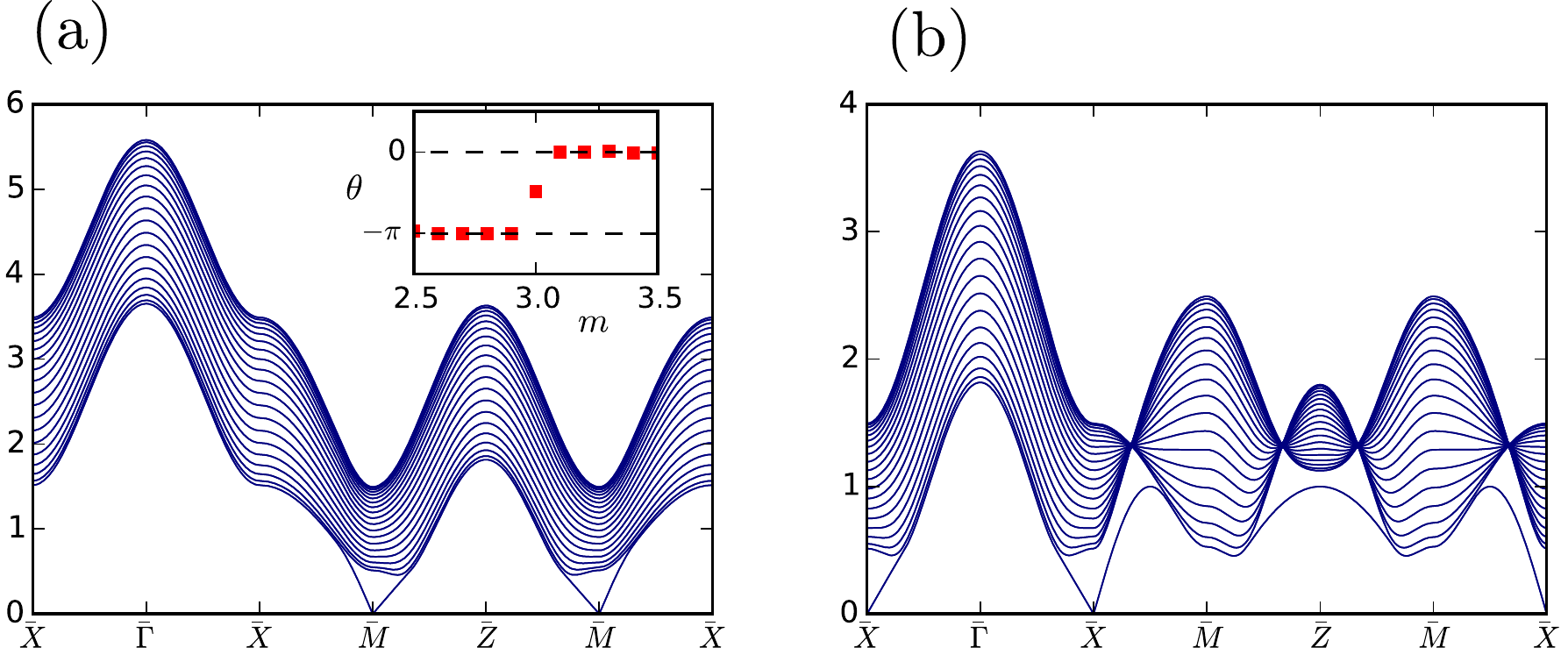}
\caption{Band structures in slab geometry for glide protected class D topological superconductor. We use $H^{DG}$ with $t_{\mu} = -1$. (a) with $m = 2.5$ the surface Majorna cone is at $\bar{M}$, while (b) with $m = 0.5$ at $\bar{X}$. The left and right moving branches are distinguished by their glide eigenvalues on the invariant lines and cannot gap out. Inset: Evolution of $\theta$ while tuning across the transition from the topological ($m<3$) to the trivial ($m>3$) phase from numerics. Note that at $m=3$ the gap closes and $\theta$ takes on an intermediate value not allowed in a gapped system. The error bars for the Monte Carlo results are smaller than the symbols.}\label{fig:ClassD}
\end{center}
\end{figure}

For appropriate choice of parameters (see Fig.~\ref{fig:ClassD}) the model is in its topological phase with gapless excitations on symmetry preserving surfaces and numerical evaluation confirms that $\theta = \pi$ (Appendix \ref{CS}). Fig.~\ref{fig:ClassD} (a) and (b) shows surface spectra with single Majorana cones pinned by particle-hole symmetry to different high symmetry points. As will be explained later, in fact the Majorana cone can only appear at $\bar{X}$ or $\bar{M}$, but not $\bar{\Gamma}$ or $\bar{Z}$ in the surface BZ (shown Fig.~\ref{glideBZ0} (b)). This is an important difference compared to the insulator case.

\subsection{Lower dimensional topological invariants}

The structure of the surface modes in the particular case of class D can be further understood from the presence of lower dimensional topological invariants (known as weak invariants or indices) associated to glide-invariant lines and planes in the Brillouin Zone. While our model has been chosen such that all 1d and 2d invariants associated to particle-hole symmetry are trivial, the presence of a mirror or glide enables new lower dimensional invariants. 

First we review the case of mirror symmetry\cite{Morimoto,Chiu}. The classification for TCSCs in class D depends on the square of mirror operator i.e. $M_{\pm}^2=\pm\id$. In particular, no nontrivial TCSCs exist with $M_-$, this means, while $\theta$ is quantized, only $\theta=0$ is allowed by symmetry. On the other hand, with the choice of $M_+$, we find a rich structure with both 1d and 2d weak invariants. Similarly to class A, there is an integer-valued 2d index (mirror Chern number) in mirror-invariant planes. Besides, there are 1d mirror $\mathbbm{Z}_2$ invariants along the high symmetry bulk lines in $y$ direction (blue lines in Fig.~\ref{glideBZ0}~(a)). These mirror $\mathbbm{Z}_2$ invariants guarantee the presence of a pair of surface zero modes of opposite parity at the corresponding surface high-symmetry momentum, which split for any other momenta generating a Majorana cone. Therefore, these 1d mirror indices determine the parity of the number of surface Majorana cones on the surface projections of the mirror invariant lines (Fig.~\ref{glideBZ0}~(b)). This leads to a $\mathbbm{Z}_2^4$ index in the case of $M_+$ mirror symmetry, as long as translation symmetry is preserved.

The case of a glide can be understood as a combination of the above two cases as we argue below. Mirror Chern numbers cannot be defined with a glide for the same reason that applies in class A. However, the smaller symmetry group with glide symmetry poses weaker constraints on the band structure compared to mirror $M_-$, allowing $\theta = \pi$ as illustrated by our model. 1d indices can still be defined, but the square of the glide operator is different for high symmetry points since it changes with $k_x$, unlike for a mirror. For the lines at $k_x=0$ we have $G_{0}^2 = -\id$, but for $k_x=\pi$, $G_{\pi}^2 = \id$. This difference is key because for $G_{0}^2 = -\id$, no mirror index exists\cite{Morimoto,Chiu}, while for $G_{\pi}^2 = \id$ there is a $\bbZ_2$ index at each high symmetry point. Therefore, surface Majorana fermion can only be found at high-symmetry points with $k_x = \pi$, i.e. the $\bar{X}$ or $\bar{M}$ points. Two cases with a single Majorana cone at $\bar{X}$ or $\bar{M}$ are realized in our proposed model for $m=0.5$ and $m=2.5$, as seen in Fig.~\ref{fig:ClassD}, in both cases $\theta = \pi$ as it is determined by the parity of the total number of Majoranas.

\subsection{Surface Dirac model}

An alternative approach to demonstrate the protection of a topological phase by a symmetry is to consider how symmetries are implemented in a generic surface theory. For example, for a regular topological insulator, the presence of time-reversal symmetry protects a single Dirac cone to be gapless. If a single Dirac cone is found at the surface, it cannot be removed until the symmetry is broken or the bulk gap closes. Two surface Dirac cones can however be gapped without breaking the symmetry.

We consider how the glide symmetry is implemented in a generic Dirac Hamiltonian at a high-symmetry point, offering an alternative explanation of the results in the previous section. As mentioned earlier, as a result of particle-hole symmetry a single Majorana cone can only appear at a high symmetry surface momentum, we will discuss the case of more Majoranas later. In a 2-band model $G_0$ that squares to $-\id$ can be chosen $i \sigma_i$ for $i = x,y,z$, but since $\sigma_x$ and $\sigma_z$ behave the same under complex conjugation we only need to consider $\sigma_x$ or $\sigma_y$. Charge conjugation is represented as $\cC = U_{\cC} \cK$ with real unitary $U_{\cC}$. A 2-band gapless Dirac Hamiltonian has the form
\eqn{
H = k_x \Gamma_x + k_z \Gamma_z
}
where the $\Gamma$'s are hermitian, anticommuting, with $\pm 1$ eigenvalues and $[\Gamma_x, G_0] = \{\Gamma_z, G_0\} = [\Gamma_i, \cC] = 0$. If we choose $G_0 = i\sigma_y$, $[U_{\cC} , G_0] = 0$, so $U_{\cC} = \id$, $\Gamma_x$ cannot be chosen to satisfy all commutation and anticommutation relations. Similarly if $G_0 = i\sigma_x$, $\{U_{\cC} , G_0\} = 0$, $U_{\cC} = \sigma_z$ and again no  $\Gamma_x$ is allowed. Therefore it is impossible to write a Dirac Hamiltonian with particle-hole symmetry and $G_{0}^2=-\id$. This shows that a single surface Majorana cone is forbidden at the $\bar{\Gamma}$ and $\bar{Z}$ points of the surface BZ (see Fig. \ref{glideBZ0} (b)) in the presence of a glide. 

On the other hand with $G_{\pi}^2=+\id$ we can choose $G_{\pi} = \sigma_x$, $U_{\cC} = \id$, $\Gamma_x = \sigma_x$ and $\Gamma_z = \sigma_z$. Now a single Majorana cone is allowed, but it cannot be gapped out, for that we would need a mass term $m\Gamma_0$ such that $\{\Gamma_0,\Gamma_i\} = \{\Gamma_0,\cC\} = [\Gamma_0, G_{\pi}] = 0$. One can check that a mass term is not allowed for any valid choice of a $2\times 2$ representation, single Majorana cones are allowed and protected at $\bar{X}$ and $\bar{M}$.

Finally, we may also consider a system with a pair of cones at opposite surface momenta. Similarly to class A\cite{FuGlide}, a pair of surface Majorana cones with different glide eigenvalues at one high symmetry point are locally protected, but can symmetrically move around the BZ and gap each other out at another point where their eigenvalues are the same. This shows that only the number of cones modulo 2 at each of $\bar{X}$ and $\bar{M}$ is stable against symmetry-preserving perturbations. The classification with full translation invariance is thus $\bbZ_2^2$, while allowing terms doubling the unit cell in the $z$ direction reduces the classification to $\bbZ_2$ counting the parity of the total number of surface Majorana cones. Such a $\bbZ_2$ index is given by the Chern-Simons 3-form in Eqn. (\ref{eqn:CS3}).

\section{Class C superconductor with glide plane in 3d}\label{sec:classC}

In this last section we consider how a singlet superconductor with SU(2) spin rotational symmetry in three dimensions may also have a topological phase. This type of superconductor belongs to class C. After appropriate rearrangement of the degrees of freedom (see Appendix \ref{BdG}) the Hamiltonian can be block-diagonalized in the spin-$S^z$ basis where the two blocks are unitarily related with identical spectra and topological properties. In the reduced problem charge conjugation is combined with a spin rotation acting as $\cC = \tau_y \cK$, with $\cC^2= - \id$, which is the main difference compared to class D.

To understand the emergence of the topological phase protected by a glide in class C, it is instructive to first consider the more familiar case of an SU(2) invariant superconductor with time-reversal symmetry, which belongs to class CI, and the anomalous response of the surface after breaking time reversal symmetry. After this, we argue how the $\theta$ term is also quantized in a class C superconductor with a glide, present an explicit model for this, and also argue how the protection of the surface states can be seen directly from the surface theory.

\subsection{Bulk invariant and SU(2) axion term}

In class CI, topological superconductors are characterized by a topological invariant $\nu$, which counts the number of pairs of surface Majorana cones. The SU(2) spin rotation symmetry allows the definition of spin Hall conductance\cite{SenthilSQH}. Once the surface is gapped by breaking time-reversal symmetry (but not SU(2) spin rotational symmetry), the surface spin quantum Hall conductance is given by\cite{Lu}
\eqn{
\sigma^S_{xy} = \frac{(\hbar/2)^2}{h} \frac{\nu}{2}
}
where $\nu$ is an even integer. Note that when $\nu=2$ the above $\sigma^S_{xy}$ is only half of that of a $d+{\text i} d$ singlet superconductor in 2d\cite{SenthilSQH}. Therefore this anomalous half-integer surface spin quantum Hall conductance serves as a probe to characterize the nontrivial 3d phase. It should be noted that the thermal Hall conductance $\kappa_{xy}$ is well defined, but it is not sufficient to characterize the topological phase.

This half-integer response can be related to an analog SU(2) bulk $\theta$ term, in an analogous way to a 3d topological insulator. To see this, we first consider the effective SU(2) continuum gauge theory that describes the spin quantum Hall superconductor in 2+1 dimensions, which captures the response of the SU(2) spin rotation invariant system coupled to a SU(2) gauge field. While this gauge field is fictitious, this treatment is useful to derive the response to an external Zeeman field\cite{ReadGreen}. This system is described by the effective action\cite{ReadGreen}
\eqn{
S = \frac{1}{4\pi} \frac{(\hbar/2)^2}{\hbar} C\int A^z \wedge \dee A^z
}
where $A^z$ is the $z$ component (in spin space) of the SU(2) gauge field, which we identify as the $z$ component of an external Zeemann field $A^z = B^z$. $C = \int \cF \in 2\bbZ$ is the Chern-number of the negative energy bands in one spin sector, which is an even integer in class C. This is analogous to the quantum Hall effect, but as we are interested in spin currents, the electric charge $e$ is replaced by $\hbar/2$ in the coupling (we take the $g$-factor $g = 1$). The spin current is
\eqn{
J^i = \delta S/ \delta A^z_i = \frac{(\hbar/2)^2}{h} C \epsilon_{ij} \partial_j B^z ,
}
so the spin Hall conductance is $\sigma^s_{xy} = \frac{(\hbar/2)^2}{h} C$. To get a proper SU(2) gauge theory, we promote $A$ to a nonabelian SU(2) gauge field by $A = \sigma_i A_{\mu}^i$, the action compatible with the previous one is given by the nonabelian Chern-Simons 3-form
\eqn{
S = \frac{1}{4\pi} \frac{(\hbar/2)^2}{\hbar} \frac{C}{2} \int \Tr \pars{ A\wedge \dee A - \frac{2i}{3} A \wedge A \wedge A},
}
where an extra $1/2$ appears to compensate for the trace. As we see, compared to the U(1) gauge theory, the coefficient is only half of the Chern number for one spin sector.

In analogy to Eq.~(\ref{theta1}), an SU(2) axion action can be defined in 3+1 dimensions in the following way:
\eqn{
S = \frac{1}{4 \pi} \frac{(\hbar/2)^2}{\hbar} \frac{\theta}{2}  \int \Tr F \wedge F
}
where $F$ is the nonabelian field strength tensor and $\theta$ is the Berry Chern-Simons 3-form (Eq.~(\ref{eqn:CS3})) for the occupied bands in one spin species. The ambiguity in $\theta$ is now $4\pi$ because of the extra factor of $1/2$. One can also check that a spatial domain wall of $4\pi$ in theta gives rise to a surface with the minimum allowed value of the spin Hall conductance in a 2d system, corresponding to $C=2$ in one spin sector. As we argued in class A, if the system has a symmetry on average that flips an odd number of spacetime dimensions, $\theta$ is quantized to $0$ or $2\pi$ modulo $4\pi$, leading to a $2\bbZ_2$ classification. The reason why the ambiguity in $\theta$ calculated from the band structure changes to $4\pi$ is quite subtle\cite{ReadGreen}.  It comes from the gauge fixing requirement that the band structure can be continuously deformed to the trivial band structure in the trivial gauge without breaking $\cC$. We refer the interested reader to Appendix \ref{CS} for details.

\subsection{Microscopic model with glide plane}

Following the same logic as for classes A and D, we now consider how the presence of a glide symmetry can protect an SU(2) invariant topological superconductor in the absence of time-reversal symmetry, i.e. in class C. As in the previous classes, the presence of an orientation reversing symmetry is sufficient to guarantee the quantization of the $\theta$ term (now to either 0 or $2\pi$), which is the topological invariant that characterizes the phase. This phase has formally the same properties as a CI topological superconductor with $\nu=2$, namely a pair of Majorana surface cones and a half-quantized spin Hall conductance upon breaking the glide symmetry on the surface.

To show this, we now consider a microscopic model in class C with a glide symmetry, demonstrating the presence of protected surface modes and computing the value of $\theta$ explicitly. In the original full Hilbert space the natural representation of the glide is $G_{\bk} = i e^{i k_x/2} \rho_x \tau_z \sigma_z $, but because of the full SU(2) spin symmetry we can cancel the spin rotation part by attaching $-i\tau_z\sigma_z$ to our definition, so we may use $G_{\bk} = e^{i k_x/2} \rho_x$. For this operator $G_{\bk}^2 = +e^{i k_x} \id$, showing the ``spinless" nature of the problem. A Hamiltonian for one spin component with these symmetries can be constructed as
\begin{align}
H^{CG}_{\bk} =& \Delta_{xy} \sin k_y \sin \frac{k_x}{2} \rho_x \tau_x + \Delta_{xz} \sin k_x \sin k_z \rho_z \tau_x + \nonumber\\
  &+ \Delta_0 (\cos k_x -\cos k_y + \alpha) \tau_y +  m_{\bk} \tau_z.
\label{eqn:HCG}
\end{align}

\begin{figure}[h]
\begin{center}
\includegraphics[width=8.5cm]{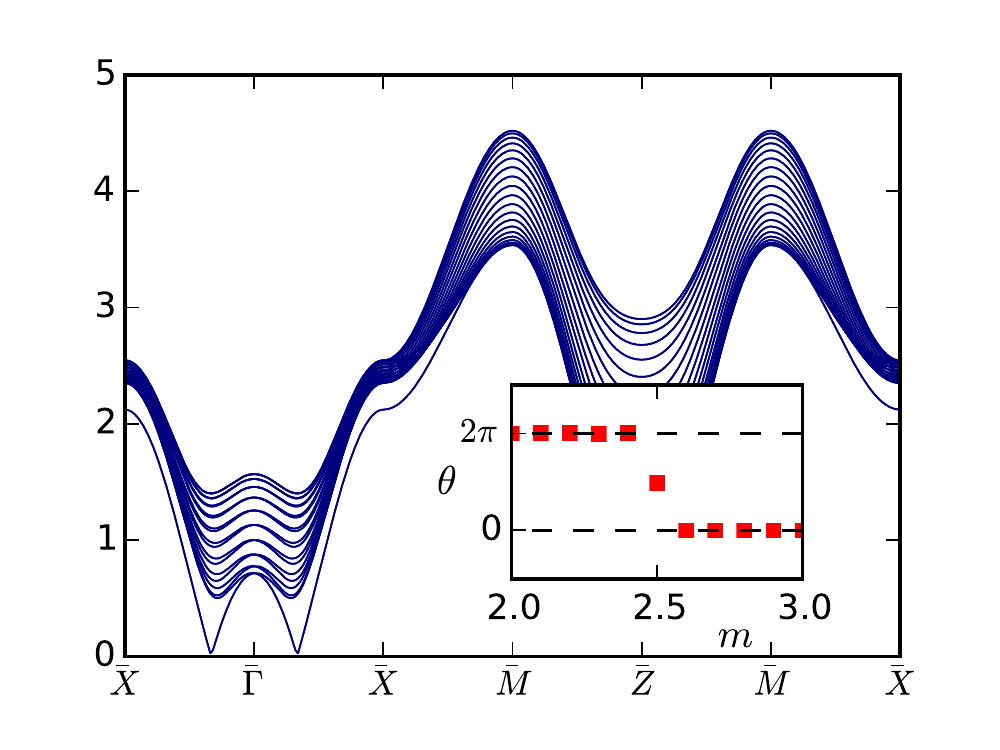}
\caption{Band structure in slab geometry for glide protected class C topological superconductor. We use $H^{CG}$ with $\Delta_{xy} = \Delta_{xz} = \Delta_0 = 1$, $\alpha = 0.5$ and $m=1.5$. Note the symmetric pair of surface Majorana cones at a generic momentum, the left and right moving branches are protected from gapping out by different glide eigenvalues. This is the spectrum for one spin sector, for the full system there is an additional spin degeneracy. Inset: Evolution of $\theta$ while tuning across the transition from the topological to the trivial phase from numerics. Note that at $m=2.5$ the gap closes and $\theta$ takes on an intermediate value not allowed in a gapped system. The error bars for the Monte Carlo results are smaller than the symbols.}\label{fig:ClassC}
\end{center}
\end{figure}

For appropriate choice of parameters (Fig.~\ref{fig:ClassC}) this Hamiltonian realizes a nontrivial topological phase with an odd number of pairs of surface Majorana cones (not counting the spin degeneracy). Class C is similar to class A as it has a $2\bbZ$ index in 2d without symmetry, corresponding to an even total Chern number in $xy$ cuts which vanishes in our model and we restrict our further discussion to this case. Again, no integer-valued mirror Chern number can be defined in glide-invariant planes, but a $2\bbZ_2$ index still survives due to the bulk quantization of $\theta$, which we have computed explicitly (Fig.~\ref{fig:ClassC} inset).

\subsection{Surface Dirac model}

The fact that an odd number of pairs of Majorana cones is protected in class C with a glide can be shown explicitly by considering the surface Hamiltonian of a pair of Majorana cones with $\cC^2=-\id$. It is instructive to show first the known case of class CI. The Hamiltonian in the vicinity of a high symmetry surface momentum (Fig. \ref{glideBZ0} (b)) is
\begin{equation}
H = \sigma_x k_x + \sigma_z k_z
\end{equation}
where the Dirac matrices are $4\times 4$ and spanned by $\sigma_i$, $\tau_j$, the identity is implicit. The particle-hole operator is $\cC = i\tau_y \cK$ and the time-reversal operator is $\cT = \sigma_y \tau_y \cK$ with $\cK$ complex conjugation. There are four possible mass terms for this Hamiltonian, $\sigma_y\tau_x, \sigma_y\tau_y, \sigma_y\tau_z,\sigma_y$. The first three masses are forbidden by $\cC$, and the last one, $\sigma_y$, is forbidden by $\cT$. Therefore, a single pair of Majorana cones cannot be gapped out at the surface in class CI.

In class C, when time reversal symmetry is broken, both a mirror or a glide can still protect the presence of a single pair Majorana cones, because as we now show, both of these symmetries forbid the $\sigma_y$ mass as well. In the presence of a mirror, we consider a reflection $z\rightarrow -z$, with operator $M_\pm$ that satisfies $M^2_\pm = \pm \id$. In the presence of a glide, we would have $G^2_{\bk} = e^{ik_x}$, but since the Majorana cones must be at opposite momenta due to $\cC$, their annihilation can only occur at $k_x=0,\pi$ and we only need to consider these two cases with $G_{0}^2 = + \id$ and $G_{\pi}^2 = -\id$, same as the situation of regular mirrors. We now discuss the two cases of $M_{\pm}$, which apply to both mirror and glide.

In the $M_+$ case, for the two eigenstates of $\sigma_x$ with positive eigenvalues, the mirror eigenvalues are the same and equal to 1, while the other two are equal to -1, as would happen if $M$ were the glide operator at $k=0$. The operator doing this is simply $M_+ = \sigma_x$. In the $M_-$ case, for the two eigenstates of $\sigma_x$ with positive eigenvalues, the mirror eigenvalues are $\pm i$ , as would happen if $M$ were the glide operator at $k=\pi$. This is achieved with $M_- = i \sigma_x \tau_z$. Note both $M_\pm$ anticommute with $\sigma_z$ because $M$ reverses $k_z$. Also note both $M_\pm$ commute with $\cC$ as we want.

Once we have the operators for $M_{\pm}$, it is easy to see that both operators forbid the mass $\sigma_y$. Since the other three masses are already forbidden by $\cC$, a pair of Majorana cones is protected in the presence of $\cC^2=-\id$ and the additional mirror/glide symmetry $M_{\pm}$.

\section{Conclusions}

In this work we classified 3d topological insulators and superconductors protected by non-symmorphic glide symmetry in classes A, D and C and presented lattice models for these phases. Our results, however, are more general. As our arguments only rely on symmetries of the effective long-wavelength response theory, the $\bbZ_2$ classification is also robust if mirror symmetry, or any symmetry reversing an odd number of spatial coordinates, is preserved on average. The cases with glide symmetry illustrate that these macroscopic considerations are insensitive to the fractional translation that accompanies the mirror operation, and identify the most robust topological invariants that are also defined (among others) with simple mirror symmetry. Glide symmetry is present in over a hundred of the 230 crystallographic space groups, and all but the 65 chiral groups contain orientation-reversing operations, so our results should be widely applicable to experimentally and numerically detect topological crystalline insulators and superconductors without time-reversal symmetry.

\section{Acknowledgements}

The authors are grateful to Joel Moore, Philipp Dumitrescu and Takahiro Morimoto for helpful conversations. This work is supported by NSF Grant No. DMR-1206515 (D.V., F. de J.), Office of Science, Basic Energy Sciences, Materials Sciences and Engineering Division, Grant No. DE-AC02-05CH11231 (F. de J.) and startup fund at Ohio State University (Y.-M. L.).

\bibliography{nonsymmorphic}

\appendix

\section{Evaluation of Chern-Simons 3-form and second Chern-form}
\label{CS}

\begin{figure}[h]
\begin{center}
\includegraphics[width=8.5cm]{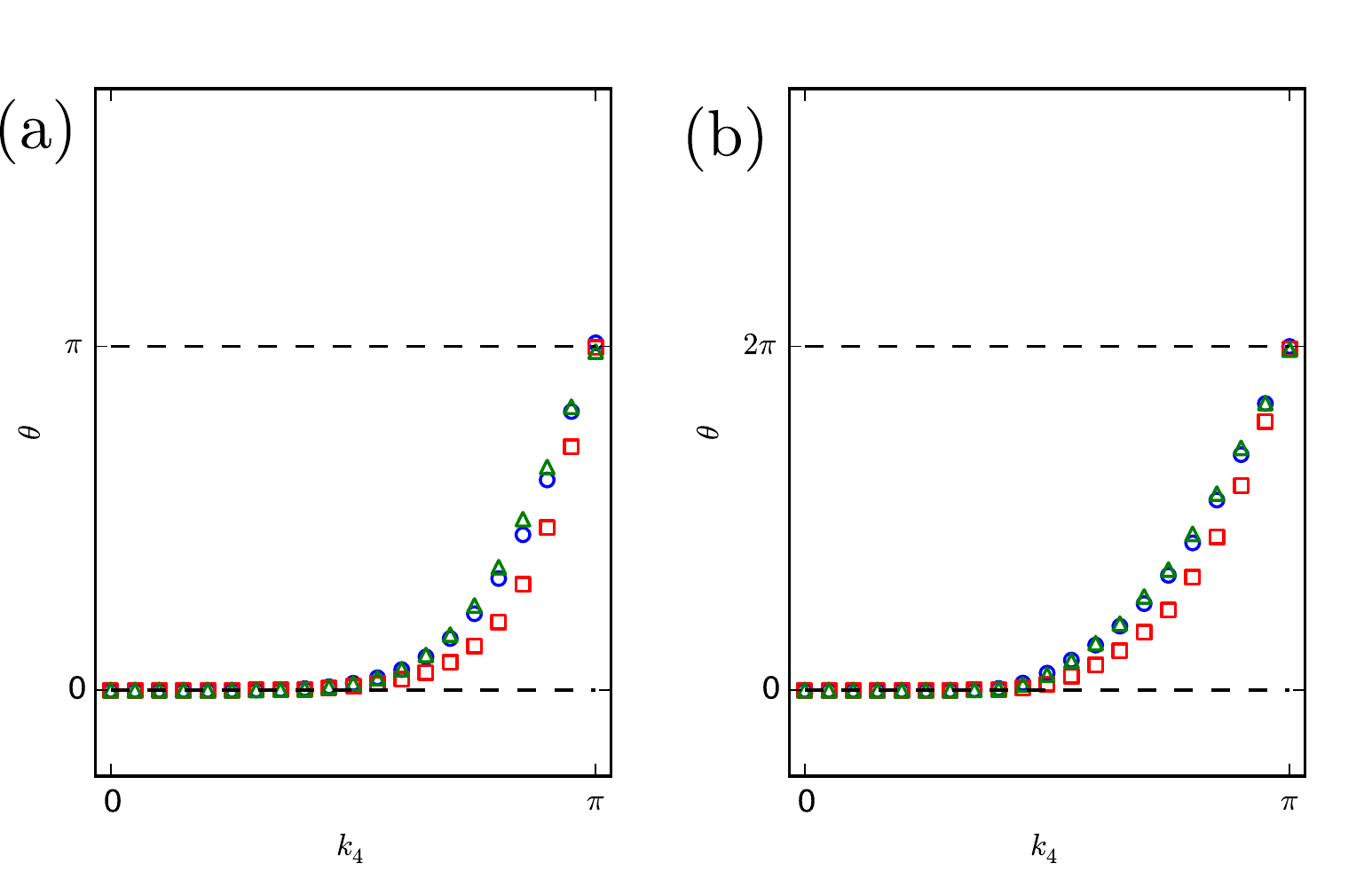}
\caption{(a) Evolution of $\theta$ during the gapped deformation $H^{AG}_{\sfk}$ from the trivial to the topological phase while breaking glide symmetry in class A. We use parameters $t_{\mu} = 1$, $\phi = 0.4$ and $m = 2$ (circles) or $m = 2.5$ (squares) and we also show a deformation with $m = 2$ but with the symmetry allowed perturbation $\beta \tau_y$ with $\beta = 0.5$ added to $H^{AG}_{\bk}$ (triangles). (b) Same plot for class C using $H^{CG}_{\sfk}$ with $\Delta_{xy} = \Delta_{xz} = \Delta_0 = 1$, $\alpha = 0.5$ and $m=1.5$ (circles) or $m = 2$ (squares) and $m = 1.5$ with symmetry allowed perturbation $\beta  \tau_x$ with $\beta = 0.5$ added to $H^{CG}_{\bk}$. The error bars are smaller than the symbols. }
\label{fig:thetasns}
\end{center}
\end{figure}

In this section we review the calculation of the magnetoelectric coupling. All our formalism is in $\bk$-space, $\dee$ denotes the exterior derivative and $\wedge$ the wedge product\cite{Nakahara}, these are antisymmetrized in the spatial indices ($\mu$, $\nu$, \textellipsis). Operators either act on the occupied band space and trace is taken over occupied band indices ($n$, $m$, \textellipsis) or the full Hilbert space of the unit cell and we supress most indices for the sake of a compact notation. The magnetoelectric coupling is defined in terms of the Chern-Simons 3-form:
\eqn{
\theta = \frac{1}{4 \pi} \int_{BZ} \Tr\pars{ \cA \wedge \dee \cA -\frac{2i}{3} \cA \wedge \cA \wedge \cA }.
}
The main difficulty about evaluating this expression is that one has to find a (patchwise) smooth gauge in the occupied band space, which is a complicated task in numerical studies. To circumvent this, we instead find a gapped deformation to a trivial state with constant Hamiltonian using a tuning parameter $k_4$, such that $\theta(k_4 = 0) = 0$, and calculate the change in $\theta$ by the 4 dimensional second Chern form\cite{Qi2008,Essin2009} that is locally gauge invariant:
\eqn{
\theta(\pi) - \theta(0) = \frac{1}{4 \pi} \int_{BZ}\int_{k_4=0}^{\pi} \Tr\pars{ \cF \wedge \cF }
}
where $\cF = \dee \cA - i \cA \wedge \cA$, the nonabelian Berry curvature in the 4 dimensional space spanned by ${\sfk_{\mu} = \pars{\bk,k_4}}$. (Note that in the convention we use $\cF\wedge\cF = \frac{1}{4} \epsilon_{\mu\nu\gamma\lambda} \cF_{\mu\nu}\cF_{\gamma\lambda}$.) We realize that $\cF$ can be written in a gauge invariant form as $\cF = i \cP (\dee \cP) \wedge (\dee \cP) \cP$ using the projector onto occupied bands $\cP_{\sfk} = \sum_{n\in occ.} \ket{u_{\sfk}^n}\bra{u_{\sfk}^n}$, which is also gauge invariant. In this formulation $\cF$ is an operator acting on the full Hilbert-space of the unit cell, but only nonzero on occupied bands, so we can extend the trace to the full Hilbert-space without changing the result. The usual components can be obtained as matrix elements between occupied states in a given basis, $\cF_{\sfk}^{nm} = \bra{u_{\sfk}^n} \cF_{\sfk} \ket{u_{\sfk}^m}$. So we arrive at the locally gauge-invariant expression
\begin{align}
&\theta(\pi) - \theta(0) = \\
&= - \frac{1}{4 \pi} \int_{BZ}\int_{k_4} \Tr\pars{\cP (\dee \cP) \wedge (\dee \cP) \cP \wedge (\dee \cP) \wedge (\dee \cP)}\nonumber
\label{eqn:CS2P}
\end{align}
that we numerically evaluate using adaptive Monte Carlo integration\cite{LepageVegas}. In order to get physical result, one must use the Bloch formalism where the orbital positions are taken into account (Appendix~\ref{conv} and \ref{sec:trfproof}), the interpolating Hamiltonians in classes A and D can be written as
\begin{align}
H^{AG}_{\sfk} =& \frac{1}{2} (1-\cos k_4) H^{AG}_{\bk} +\nonumber\\
 &+ \sin k_4 \sin(k_x/2) \rho_y + \frac{1}{2} (1 + \cos k_4) \tau_z\\
H^{DG}_{\sfk} =& \frac{1}{2} (1-\cos k_4) H^{DG}_{\bk} +\nonumber\\
 &+ \sin k_4 \sin(k_x/2) \rho_x \tau_x + \frac{1}{2} (1 + \cos k_4) \tau_z
\end{align}
with $H^{AG}_{\bk}$ and $H^{DG}_{\bk}$ given in (\ref{eqn:HAG}) and (\ref{eqn:HDG}) respectively. We find that $\theta$ is quantized to $0$ or $\pi \pmod{2\pi}$ to high accuracy, a result that is robust against symmetry allowed perturbations of the final Hamiltonian and deformations of the interpolation as long as the bandgap does not close (Fig.~\ref{fig:thetasns} (a)).

In class C we have to reconcile the change of ambiguity in $\theta$ from $2\pi$ to $4\pi$ from the band structure point of view. A natural gauge choice for a class C system is the requirement that $\left| u_{\bk}^o \right\rangle = i \tau_y \left| u_{-\bk}^u \right\rangle^*$ that relates unoccupied states at $\bk$ with occupied states at $-\bk$. Gauge transformations preserving this condition satisfy $\tau_y U^o_{\bk} = (U^u_{-\bk})^* \tau_y$ where $U^{o/u}$ act on the occupied/unoccupied bands. This constraint is not sufficient to remove the $2\pi$ gauge ambiguity in $\theta$ coming from the winding number of $U^o_{\bk}$, as any $U^o_{\bk}$ is allowed as long as it is accompanied by the appropriate $U^u_{\bk}$.

So how can we define a proper bulk invariant? The idea is to prove that for a cyclic gapped deformation of the band structure (with $k_4$ as the tuning parameter) the second Chern form
\begin{equation}
\frac{1}{4 \pi} \int \Tr \cF \wedge \cF
\end{equation}
is quantized to multiples of $4\pi$ as long as particle-hole symmetry in any 3d cut is preserved (as opposed to multiples of $2\pi$ without symmetry). This is proved in Ref. \onlinecite{TeoKaneDefect}, for our case $D=3$ is the dimension of $k$-space and $\delta = 1$ the dimension of real space, for us this is $k_4$, the tuning parameter that is not affected by particle-hole symmetry. The result is, for $D-\delta = 2$ in class C the classification is $2 \mathbb{Z}$, proving our conjecture.

Now, if we find a gapped particle-hole symmetric deformation from the trivial band structure to the glide symmetric one, we can calculate the difference $\Delta \theta = \theta(\pi) - \theta(0)$ in the 3d Chern-Simons forms between the initial and final states using the locally gauge invariant expression of the second Chern form in terms of the band projector. $2\pi$ will be different from zero, as the ambiguity introduced by different deformations is $4\pi$. Such deformation to the trivial state always exists, as there are no nontrivial phases in 3d class C without any other symmetry. Note that the formula in terms of $\cF$ requires a continuous gauge choice throughout the deformation and $\theta(0)$ is only zero in the trivial gauge. We can view this as a nontrivial gauge fixing condition demanding that the band structure is continuously deformable to the trivial state in the trivial gauge, in this gauge the Chern-Simons 3-form in the final state gives the same result. However, we know of no method of checking whether this condition is satisfied other than explicitelly constructing a deformation. On the other hand, the formula with the projector is completely gauge-invariant, insensitive to discontinuous large gauge changes that would add extra $2\pi$'s to the formula with $\cF$, so we do not have to worry about the band structure satisfying any gauge condition using this method of computation.

For the specific model in equation (\ref{eqn:HCG}) we use the deformation preserving $\cC = \tau_y \cK$:
\begin{align}
H^{CG}_{\sfk} =& \frac{1}{2} (1-\cos k_4) H^{CG}_{\bk} +\nonumber\\
 &+ \sin k_4 \sin(k_x/2) \rho_y\tau_x + \frac{1}{2} (1 + \cos k_4) \tau_z.
\end{align}

Our numerical results give $\theta = 0$ and $2\pi \pmod{4\pi}$ to a high accuracy for the trivial and topological phases respectively (Fig.~\ref{fig:thetasns} (b)).

\section{Relation to earlier definition of $\bbZ_2$ index in class A}
\label{bentBZ}

\begin{figure}[h]
\begin{center}
\includegraphics[width=8.5cm]{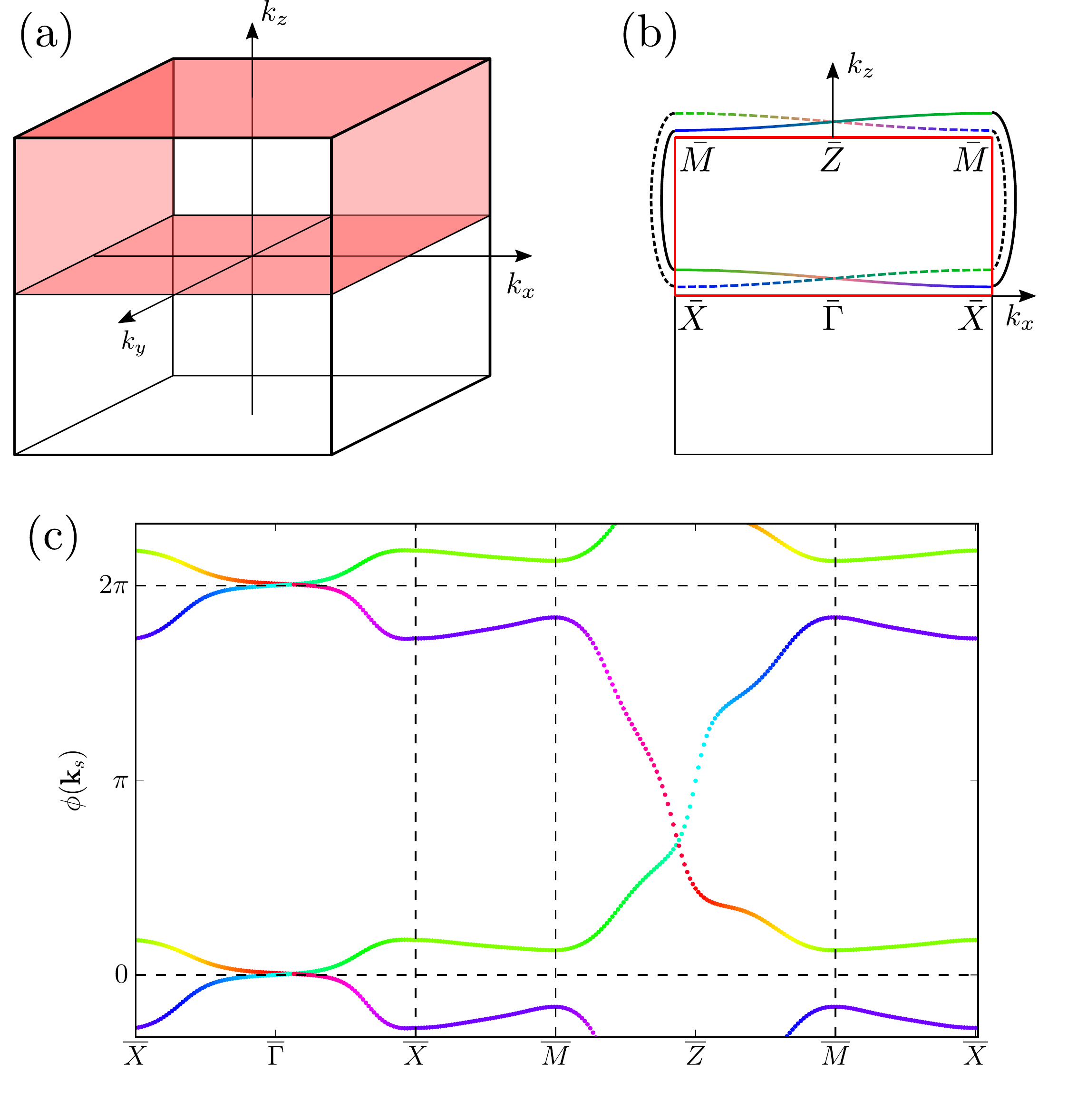}
\caption{Illustration of our definition of the glide Chern number. (a) Bulk BZ of a glide symmetric crystal with a possible choice of the ``bent" BZ in red (shaded). The glide Chern number is calculated by integrating the Berry flux following a single band around this torus. (b) Surface BZ for a cut normal to $y$ with projection of the ``bent" BZ in red (thick) and a sketch of an occupied band pair along this loop. On the invariant planes ($k_z = 0$ and $k_z = \pi$) the glide eigenvalue (color code) distinguishes the two bands. On the non-symmetric part (other values of $k_z$) bands are generically nondegenerate. The glide Chern-number is calculated by counting the winding number of $\phi(\bk_s)$ following a single band (solid or dashed) around the $\bar{X}-\bar{\Gamma}-\bar{X}-\bar{M}-\bar{Z}-\bar{M}-\bar{X}$ loop. (c) Value of $\phi(\bk_s)$ around the loop for the minimal model in the nontrivial phase. Color code indicates glide eigenvalue in the invariant planes. Note that following one band around, the phase winds $2\pi$, indicating that the glide Chern-number is odd.}\label{glideBZ}
\end{center}
\end{figure}

Here we review the alternate definition of the $\bbZ_2$ index in class A introduced in Ref. \onlinecite{FuGlide}.
There are two planes in the 3d BZ that are invariant under a mirror or glide, at $k_z=0$ and $\pi$, in the rest of the BZ glide does not act locally, bands do not have a well defined eigenvalue and symmetry allowed perturbations generically destroy most degeneracies. A $2\times 2$ Hamiltonian without any symmetry constraints only has pointlike degeneracies between occupied bands in 3d (Weyl nodes), which allows us to choose a surface in the BZ that connects the $k_z = 0$ and $\pi$ planes such that the all the bands are nondegenerate. We will choose a constant $k_x$ surface for simplicity and show later that the result is insensitive to this choice or the assumption that there are only pointlike degeneracies.

Now we can choose a surface in the bulk BZ that includes the two invariant planes plus a surface connecting them traversing half the BZ, say at $k_x = 0$ (see Fig.\ref{glideBZ} (a)). We can follow a band around this surface (Fig.\ref{glideBZ} (b)), this is possible because of the well defined glide eigenvaue in the invariant planes and the generic lack of degeneracies in the connecting surfaces. We can compute the ``glide Chern number" corresponding to one such band on the ``bent BZ"\cite{BernevigPGS2} by counting the winding number of the Berry-connection integral  integrals in the $y$ direction
\eqn{
\phi(\bk_s(t)) = \int_0^{2\pi} \cA^y (\bk) dk_y.
}
where $\cA^{\mu}(\bk) = i \bra{u_{\bk}}\partial^{\mu} \ket{u_{\bk}}$ and the curve $\bk_s(t) = (k_x(t),k_y(t))$ encloses half of the surface BZ (Fig.\ref{glideBZ} (b)). In the case when there are multiple pairs of conduction bands with  we naturally sum the Chern numbers for bands with the same glide eigenvalue. This quantity $\pmod{2}$ is a well defined topological invariant because pushing a Weyl node across one of the connecting surfaces can only change the winding by 2. A definition that does not rely on distinguishing the two bands in the nonsymmetric region is obtained realizing that the winding number $\pmod{2}$ is the same as the parity of crossings regardless of direction of an arbitrarily set branch cut by $\phi(t)$ throughout the invariant planes and one copy of the connecting plane\cite{FuGlide}. Even when the bands are degenerate, phases of eigenvalues of the nonabelian Wilson loop operators along the $y$ direction give equivalent quantities. This definition, while in principle well defined, still suffers from problems of distinguishing crossings from anticrossings in a many-band model at finite $\bk$-space resolution\cite{VanderbiltWC}.

The fact that the Berry flux through a closed surface for a set of bands that are separated from all other bands on this surface is nonzero signals the presence a Weyl-node inside the bounded region connecting this set of bands to some other bands. This Weyl-node connects valence/conduction bands among themselves, the systems we consider are fully gapped. This Weyl-node is also present in mirror symmetric insulators with different mirror Chern-numbers in the two invariant planes (sec. \ref{AM}). We can think of the 3d band structure as an interpolation of 2d systems in the $k_x k_y$ cuts as the tuning parameter $k_z$ evolves from $0$ to $\pi$. To interpolate 2d systems between the two cuts, we have to close and open a Dirac node between the occupied bands connected to different mirror sectors to transfer Berry flux, which corresponds to a Weyl node inside the half 3d BZ. This Weyl node is topologically protected and must exist somewhere in the half BZ, mirror symmetry at the high symmetry planes prevents it from locally annihilating with its opposite chirality mirror image that lives in the other half of the BZ.

\section{BdG formalism}
\label{BdG}

In this appendix we summarize general results of the BdG formalism used for superconductors\cite{Schnyder2008}, with special emphasis to representations of symmetries in class C.

The general form of a Hamiltonian for a superconductor without any additional symmetry is
\begin{align}
\mathcal{H} &= \sum_{\bk} \pars{\begin{array}{cc}
c_{\bk}\dag & c_{-\bk}
\end{array}}
\pars{\begin{array}{cc}
\epsilon_{\bk}   & \Delta_{\bk} \\
\Delta_{\bk}\dag & -\epsilon_{-\bk}^T
\end{array}}
\pars{\begin{array}{c}
c_{\bk} \\ c_{-\bk}\dag
\end{array}} = \nonumber\\
&=\sum_{\bk} \pars{\begin{array}{cc}
c_{\bk}\dag & c_{-\bk}
\end{array}}
H_{\bk}
\pars{\begin{array}{c}
c_{\bk} \\ c_{-\bk}\dag
\end{array}}
\end{align}
where $c$ is a vector formed of all the electron anihilation operators of the unit cell and the BdG Hamiltonian $H_{\bk}$ is a $2N\times 2N$ block matrix with $N$ orbitals in the unit cell. $\epsilon_{\bk} = \epsilon_{\bk}\dag$ (Hermiticity) and $\Delta_{\bk} = -\Delta_{-\bk}^T$ (Fermi statistics). It is customary to denote the Pauli matrices acting on the particle-hole space (the block structure of $H$) $\tau_{\mu}$. The paricle-hole symmetry is represented by $\mathcal{C} = \tau_x \mathcal{K}$ for usual electronic systems such that $\mathcal{C}^2 = \mathbbm{1}$. This symmetry restricts the BdG Hamiltonian as
\eqn{
\mathcal{C} H_{\bk} \mathcal{C}^{-1} = \tau_x H_{-\bk}^* \tau_x = - H_{\bk}
}
which is automatically satisfied by this form of the Hamiltonian. This transformation relates excitations with opposite energy and momentum in the doubled spectrum.

At this point we digress to discuss transformation properties of the BdG Hamiltonian under physical symmetry and gauge transformations. A general transformation on the particle-hole degrees of freedom has the form
\eqn{
\pars{\begin{array}{c}
c_{\bk} \\ c_{-\bk}\dag
\end{array}
}  \rightarrow
\pars{\begin{array}{cc}
U_{\bk}   & W_{\bk} \\
W_{-\bk}^* & U_{-\bk}^*
\end{array}}
\pars{\begin{array}{c}
c_{\bk} \\ c_{-\bk}\dag
\end{array}}
}
because we require that the transformed particle and hole-like operators are still related by hermitian conjugation. Moreover, preserving the fermionic commutation relations restricts the block matrix to be unitary.

Now we turn to the case with spin rotation symmetry. We split the $c$ vector in 2 halves for the spin $z$ component $c_{\bk} = \pars{\begin{array}{cc}c_{\bk\uparrow} & c_{\bk\downarrow}\end{array}}^T$ and rewrite the Hamiltonian in a $4N\times 4N$ block form where $N$ is the number of orbitals not counting spin. The U(1) spin rotation invariance around the $z$ axis requires $H_{4}$ to commute with the generator of the rotations $\sigma_z \tau_z$ which restricts it to the block-diagonal form
\eqn{
H_{4\bk} =
\pars{\begin{array}{cccc}
\xi_{\bk\uparrow} & 0                   & 0                     & \delta_{\bk}   \\
0                 & \xi_{\bk\downarrow} & -\delta_{-\bk}^T      & 0         \\
0                 & -\delta_{-\bk}^*    & -\xi_{-\bk\uparrow}^T & 0        \\
\delta_{\bk}\dag  & 0                   & 0                     & -\xi_{-\bk\downarrow}^T
\end{array}}
.}
Imposing spin rotation symmetry for the $x$ axis as well (commutation with $\sigma_x \tau_z$) means $\xi_{\bk\uparrow} = \xi_{\bk\downarrow} =: \xi_{\bk}$ and $\delta_{\bk} = \delta_{-\bk}^T$. If we now define a new set of operators $d_{\bk\sigma} = \pars{\begin{array}{cc}c_{\bk\sigma} & c_{-\bk\bar{\sigma}}\dag\end{array}}^T$ with well defined momentum and spin quantum number, we can rewrite the Hamiltonian as the sum of two $2N\times 2N$ terms for the two spin orientations
\eqn{
\mathcal{H} = \sum_{\bk\sigma} d_{\bk\sigma}\dag H_{2\bk\sigma} d_{\bk\sigma}
}
where
\eqn{
H_{2\bk\uparrow} = \pars{\begin{array}{cc}
\xi_{\bk}   & \delta_{\bk} \\
\delta_{\bk}\dag    & -\xi_{-\bk}^T
\end{array}}
}
and $H_{2\bk\uparrow} = r_z H_{2\bk\downarrow} r_z =: H_{2\bk}$ with Pauli matrices $r_{\mu}$ acting on the space with the two components of $d$. The unitary relation between the two spin sectors guarantees that the spectrum is doubly degenerate, for every eigenstate there is another state with the same energy and momentum but opposite spin.

The constraints on the form of $H_{2\bk}$ can be summarized as
\eqn{
\mathcal{C} H_{2\bk} \mathcal{C}^{-1} = r_y H_{2-\bk}^* r_y = - H_{2\bk}
}
where we introduced the new particle-hole conjugation operator $\mathcal{C} = r_y \mathcal{K}$ with $\mathcal{C}^2 = -\mathbbm{1}$ (in the body of the paper we use $\tau$ instead of $r$ for this set of Pauli matrices as well). This operator relates states with opposite energy and momentum but the same spin. Note that this operator differs from the original particle-hole conjugation in that it is combined with a spin flip, the physical symmetry should also reverse spin. As the symmetry is antiunitary and squares to $-\mathbbm{1}$, a zero energy eigenstate at an invariant momentum must be doubly degenerate (on top of the spin degeneracy that is always present) by the same reasoning that proves Kramers degeneracy with $\mathcal{T}^2 = -\mathbbm{1}$.

\section{Conventions for Bloch functions}
\label{conv}

There are two widely used conventions to define the Bloch basis functions. When appropriate we use the convention where we define Bloch basis functions $\ket{\widetilde{\chi}_{\bk}^a}$ in terms of the orbitals of the unit cell $\ket{\widetilde{\chi}_{\bk}^a} = \sum_{\bR} e^{i \bk \bR} \ket{\phi_{\bR}^a}$ where $\bR$ is the unit cell coordinate and $a$ the orbital index. Note the absence of phase factors corresponding to the position of the orbitals within the unit cell, so the basis functions are strictly periodic in the BZ. While in this convention the information about the position of the orbitals is lost, thus the polarizations computed via Berry vector potential integrals do not equal the true Wannier center positions, the Bloch Hamiltonian is BZ periodic, making some derivations more transparent.

In the other convention we define $\ket{\chi_{\bk}^a} = \sum_{\bR} e^{i \bk (\bR+\mathbf{r}_a)} \ket{\phi_{\bR}^a}$ where $\mathbf{r}_a$ is the position of the $a$-th orbital in the unit cell. The two conventions are related by the operator $W_{\bk}$ with $W^{ab}_{\bk} = \delta^{ab} e^{i \bk \mathbf{r}_a}$ such that $\ket{u_{\bk}^n} = W_{\bk}\ket{\widetilde{u}_{\bk}^n}$. Furthermore, the new Bloch wave functions obey the boundary condition $\ket{u_{\bk+\bG}^m} = W_{\bG} \ket{u_{\bk}^{m'}}$ and operators (including the Bloch Hamiltonian) satisfy $\mathcal{O}_{\bk+\bG} = W_{\bG} \mathcal{O}_{\bk} W_{\bG}^{-1}$ where $\bG$ is a primitive reciprocal lattice vector. This convention is usually assumed in formulae for electromagnetic response, as the naive Peierls substitution $\bk \to \bk + \mathbf{A}$ only gives the correct phase factor for hopping in this case. The two conventions give equivalent results for quantized topological indices in most symmorphic cases, provided there is a continuous, symmetry preserving deformation of the lattice, such that all the orbitals are brought to the same point in the unit cell. In nonsymmorphic lattices however, this is never possible, as the shortest orbit of a point in the unit cell under the symmetry group modulo lattice vectors is longer than one, there is no crystal with one site per unit cell obeying a nonsymmorphic symmetry. For example with a single essential nonsymmorphic symmetry translating in the $x$ direction one needs at least $n$ lattice sites that can be arranged such that the positions are $\mathbf{r}_a = \bbt_x a/ n$ for $a = 1 \textellipsis n$, so $W_{\bG_x}^{ab} = \delta^{ab} e^{2\pi i a/n}$ and $W_{\bG} = \id$ for perpendicular directions. In our glide examples this results in $W_{\bG_x} = \rho_z$, where $\rho$ acts on the space of the two sublattices.

\section{Proof of quantization of $\theta$}
\label{sec:trfproof}

In this appendix we provide a formal microscopic proof of our claim that any orientation-reversing space group (SG) operation quantizes $\theta$.

First we review the representations of space group operations in $\bk$-space. We use the convention with Bloch basis functions $\ket{\chi_{\bk}^{\bx l}} = \sum_{\bR} e^{i \bk (\bR+\bx)} \ket{\phi_{\bR+\bx}^{l}}$, where we split the orbital index $a = (\bx,l)$, $\bx = \br_a$ labels the sites of the unit cell by their real space position and $l$ is an on-site orbital index accounting for spin, orbital angular momentum, etc. (the values $l$ can take may depend on $\bx$). A useful property of this basis is that it is periodic in the real space coordinate (insensitive to the choice of the unit cell), i.e. $\ket{\chi_{\bk}^{\pars{\bx+\bR} l}} = \ket{\chi_{\bk}^{\bx l}}$ for any lattice vector $\bR$. On the other hand, this basis is not BZ-periodic, instead $\bk$ and $\bk+\bG$ are related by a constant transformation as explained in Appendix \ref{conv}. We emphasize that our treatment is not specific to tight-binding models, the same can be told in the continuum, there $\bx$ is the continuous index for position in the unit cell and $l$ stands for the spin only. To go to the tight-binding approximation, we restrict the Hilbert-space to a finite set of orbitals per unit cell, the only assumption we make is that there exists a basis of localized states such that orbitals centered on different sites span orthogonal subspaces\footnote{We also assume that a lattice-periodic gauge-field configuration exists, in electronic insulators it suffices to demand vanishing net magnetic flux through any face of the unit cell to ensure this.}.

Consider a general SG operation $g = \brac{O \middle| \bt}$ ($O$ is an orthogonal rotation and $\bt$ a translation) acting on one of the basis states
\eqn{
g \ket{\phi_{\bR + \bx}^{l}} = U_{\bx}^{l'l} \ket{\phi_{g\pars{\bR + \bx}}^{l'}} =
U_{\bx}^{l'l} \ket{\phi_{O\pars{\bR + \bx}+\bt}^{l'}}
}
where $U$ is the site and $g$-dependent unitary representation on the local orbitals, a double representation if the model is spinful. Applying this to the Bloch basis functions, with simple algebra we find
\eqn{
g \ket{\chi_{\bk}^{\bx l}} = e^{-i\pars{O\bk}\bt} U_{\bx}^{l'l} \ket{\chi_{O\bk}^{g\bx, l'}}
}
with $g\bx = O\bx + \bt$ that is understood as a permutation of sites at the same Wyckoff position. Grouping indices back together, this can be written as $g \ket{\chi_{\bk}^{a}} = e^{-i\pars{O\bk}\bt} U^{ba} \ket{\chi_{O\bk}^{b}}$ with $U^{\pars{\bx,l},\pars{\bx',l'}} = U^{l,l'}_{\bx} \delta_{\bx',g\bx}$.

The key observation is that in this basis the $\bk$-dependence decouples as a single phase factor. Consider the transformation of a Bloch eigenstate in the $n$-th band $\ket{u^n_{\bk}} = n^a_{\bk} \ket{\chi_{\bk}^{a}}$. The symmetry transformation results in a state at $O\bk$, the coefficients transform as $(gn)_{O\bk}^a = e^{-i(O\bk)\bt} U^{ab} n_{\bk}^b$ or in a compact notation $g \pars{n_{\bk}} = (gn)_{O\bk} = e^{-i\pars{O\bk}\bt} U n_{\bk}$. As $g$ is a symmetry operation, the transformed state is again an eigenstate of the Bloch Hamiltonian with the same energy, but at $O\bk$. As a consequence, the transformation of occupied band projector operator $\cP_{\bk} = \sum_{n\in\textnormal{occ.}} n_{\bk} n_{\bk}\dag$ reads
\eqn{
(g\cP)_{O\bk} = \sum_{n\in\textnormal{occ.}} (gn)_{O\bk}(gn)_{O\bk}\dag = U \cP_{\bk} U\dag.
}
So if $g$ is a symmetry, such that $(g\cP)_{\bk} = \cP_{\bk}$, any gauge invariant quantity that can be expressed through $\cP_{\bk}$ are invariant if the $\bk$-space coordinates are also transformed.

Now we are in a position to prove that $\theta$ is invariant under orientation preserving SG operations and changes sign (thus becomes quantized) under orientation reversing ones. First consider a symmorphic SG operation $g = \brac{O \middle| 0}$ acting as $(gn)_{O\bk} = U n_{\bk}$. Note that in this convention that reproduces the microscopic expression\cite{Essin2009} for the diagonal magnetoelectric coupling, the Berry connection is calculated using the coefficients only as $\cA^{nm}_{\bk} = i n_{\bk} \dag \dee m_{\bk}$ and we drop terms that come from the derivatives of the basis states. It transforms under $g$ as
\begin{align}
\pars{g\cA}_{\mu,\bk}^{nm} &= i \pars{gn}_{\bk}\dag \dee \pars{gm}_{\bk} = i n_{O^{-1}\bk}\dag U\dag \dee U m_{O^{-1}\bk} = \\ \nonumber
&= O_{\mu\nu} \cA^{nm}_{\nu,O^{-1}\bk}.
\end{align}
The constant $U$ cancels and the 3d BZ integral (\ref{eqn:CS3}) for $\theta$ picks up a factor of $\det O$ from the point-group rotation of $\bk$-space, so $g\theta = \pars{\det O} \theta$.

We will reduce the general case to the previous symmorphic one. Consider a continuous family of transformations represented as $(g\pars{k_4}n)_{O\bk} = e^{-i\pars{O\bk\cdot\bt} k_4} U n_{\bk}$ where $\bt$ is fixed and $0 \leq k_4 \leq 1$ is a tuning parameter, $k_4 = 1$ corresponding to the actual SG symmetry $g\pars{k_4 = 1} = g = \brac{O\middle|\bt}$. We will show that $\theta(k_4)$ calculated from a wave function transformed by $g\pars{k_4}$ is independent of $k_4$. This is to be expected, considering the special case of $U=\id$ and $O=\id$ corresponds to a mere shift of the spatial origin by $k_4 \bt$. We can calculate the change in $\theta$ as $k_4$ changes from 0 to 1 by evaluating the second Chern form $\int_{\textnormal{BZ}\times [0,1]} \cF \wedge \cF$ with $k_4$ as the fourth coordinate. Using the expression (\ref{eqn:CS2P}), as $\cP$ is $k_4$-independent, we see that $\theta$ is unchanged through this process. $g\pars{k_4 = 0}$ has the same form as a symmorphic SG operation discussed above, those considerations are still valid even though $g\pars{0}$ is not a symmetry.

The same result can be derived by direct substitution, but one encounters a subtlety we discuss now. Using the transformed wave function ${n'}_{\bk} = e^{i\bk\bt} n_{\bk}$ we find ${\cA'}_{\mu}^{nm} = \cA_{\mu}^{nm} - \bt_{\mu}\delta^{nm}$, substituting it in (\ref{eqn:CS3})
\eqn{
\theta' = \theta - \frac{1}{4\pi}\int \epsilon_{\mu\nu\lambda} \Tr \cF_{\mu\nu} \bt_{\lambda}.
}
As $\bt$ is constant the extra term can be expressed as a linear combination of Chern numbers in various cuts of the Brillouin zone. If the Chern numbers vanish, the correction is zero, or if $\bt$ is a lattice vector, it is an integer multiple of $2\pi$ that does not change the value of $\theta$ modulo $2\pi$. However, $\bt$ can be an arbitrary vector if we interpret the above transformation as a shift of the entire crystal, or equivalently as a redefinition of the spatial origin. This shows that $\theta$ is ill defined with nonzero Chern numbers, we rationalize this observation below.

This situation is analogous to the problem of polarization in 2d Chern-insulators\cite{Coh2009}. From a mathematical point of view, there the polarization becomes ill defined because it is not possible to choose a gauge where the wave functions are BZ periodic in the presence of nonzero Chern-number. Similarly, the Chern-Simons 3-form is not invariant under gauge transformations that are not periodic over the BZ, one needs to choose a periodic gauge to fix its value, but the Chern-number prevents this. Equation (\ref{eqn:CS2P}) misses the extra correction term because, while $\cP$ stays BZ periodic throughout the deformation, $\cA_4$ does not, to apply Stokes' theorem one has to add surface terms for $\partial\textnormal{BZ}\times [0,1]$, that exactly reproduces the correction. However this does not make a difference in the cases with vanishing Chern-numbers discussed in this paper. $\cA_4$ fails to be periodic because the boundary conditions for the Bloch functions change throughout the deformation. To guarantee that such surface terms do not appear during the deformation to the trivial state used to calculate $\theta$, we prescribe that these deformations should be made keeping the lattice sites (thus the boundary condition) fixed, which is always possible.

We can continue the analogy from a physical point of view. Polarization measures surface charge, but on the surface of a Chern-insulator charge is no longer conserved due to the chiral anomaly. $\theta$ measures the fractional part of the surface Hall conductance $\sigma_{xy}^S$. However, when the bulk has a Hall conductance per transverse unit cell $\sigma_{xy}/a_z$, depending on the definition of the ``surface layer'' it may contribute more or less to $\sigma_{xy}^S$. We may fix the boundary of the ``surface layer'' and push the crystal in the $z$ direction by $t_z$. The surface Hall conductance changes exactly by $t_z \sigma_{xy}/a_z$, in agreement with our formal result. We conclude that there is no natural zero for $\theta$ with nonzero Chern-number, consequently our result about the quantization of $\theta$ is only meaningful if we restrict to the case of vanishing total Chern-numbers, as we did throughout this work.

To summarize, we proved that $\theta$ transforms under a generic SG operation $g = \brac{O\middle|\bt}$ as
\eqn{
g\theta = \pars{\det O} \theta
}
meaning that $\theta$ is quantized to $0$ or $\pi$ by any orientation-reversing SG symmetry.

\end{document}